\documentclass[pdflatex,sn-mathphys-num]{sn-jnl}

\usepackage{graphicx}%
\usepackage{multirow}%
\usepackage{amsmath,amssymb,amsfonts}%
\usepackage{amsthm}%
\usepackage{mathrsfs}%
\usepackage[title]{appendix}%
\usepackage{xcolor}%
\usepackage{textcomp}%
\usepackage{manyfoot}%
\usepackage{booktabs}%
\usepackage{algorithm}%
\usepackage{algorithmicx}%
\usepackage{algpseudocode}%
\usepackage{listings}%
\usepackage{svg}
\usepackage{subcaption}
\usepackage{makecell}

\usepackage[font=small,labelfont=bf]{caption}

\theoremstyle{thmstyleone}%
\theoremstyle{thmstyletwo}%

\theoremstyle{thmstylethree}%

\begin{document}

\title[Article Title]{Can We Fix Social Media? Testing Prosocial Interventions using Generative Social Simulation}


\author*[1]{\fnm{Maik} \sur{Larooij}}\email{m.k.larooij@uva.nl}
\author*[1]{\fnm{Petter} \sur{Törnberg}}\email{p.tornberg@uva.nl}


\affil*[1]{\orgname{Institute for Logic, Language, and Computation (ILLC), University of Amsterdam}}



\abstract{Social media platforms have been widely linked to societal harms, including rising polarization and the erosion of constructive debate. Can these problems be mitigated through prosocial interventions? We address this question using a novel method -- generative social simulation -- that embeds Large Language Models within Agent-Based Models to create socially rich synthetic platforms. We create a minimal platform where agents can post, repost, and follow others. We find that the resulting following-networks reproduce three well-documented dysfunctions: (1) partisan echo chambers; (2) concentrated influence among a small elite; and (3) the amplification of polarized voices -- creating a ``social media prism'' that distorts political discourse. We test six proposed interventions, from chronological feeds to bridging algorithms, finding only modest improvements -- and in some cases, worsened outcomes. These results suggest that core dysfunctions may be rooted in the feedback between reactive engagement and network growth, raising the possibility that meaningful reform will require rethinking the foundational dynamics of platform architecture.}





\maketitle
\newpage
\section{Introduction}\label{sec1}
Political discourse in the digital age is increasingly shaped by a small number of dominant social media platforms -- systems that influence what information people encounter, whose voices are amplified, and how political conflict is perceived. Once hailed as catalysts for a revitalized public sphere \cite{benkler2006wealth,shirky2008here, dahlgren2005internet, papacharissi2002virtual}, many scholars now agree that these platforms provide limited support for the forms of constructive political dialogue deemed vital to democratic life \cite{benton2024harm, mckernan2023echo}. The platforms have been criticized for insulating users from opposing perspectives \cite{sunstein2018republic}, for concentrating visibility and influence in the hands of a small elite of users \cite{boy2023display,tornberg2022sharing}, and for amplifying sensational or divisive content, producing a distorted ``social media prism'' \cite{bail2022breaking} through which politics appears more extreme and conflictual. While the downstream consequences of these dynamics remain debated \cite{guess2023social}, a large body of research has examined possible links between social media use and polarization \cite{shmargad2020sorting, del2016echo}, radicalization \cite{Rietdijk2021-RIERPA-3,tornberg2024intimate}, and the spread of misinformation \cite{tornberg2018echo, del2016spreading, choi2020rumor}. 

Recent scholarship has called for a shift from diagnosing problems to designing platforms that actively foster \emph{prosocial} outcomes and better support constructive political discourse \cite{dorr2025research, bak2021stewardship}. Yet assessing such interventions is methodologically difficult: most studies rely on observational data, which cannot capture counterfactual scenarios such as how discourse might change under different algorithms or interface designs \cite{tucker2018social}. This challenge has intensified as major platforms have restricted researcher access, limiting APIs and other data channels \cite{freelon2018computational, bruns2021apicalypse}. 

This paper addresses this gap using a novel approach -- \emph{generative social simulation} -- that embeds Large Language Models (LLMs) within Agent-Based Models (ABMs) to create socially rich synthetic platforms \cite{park2023generative, tornberg2023simulating}. This method enables the controlled testing of interventions that cannot be implemented or observed on live platforms, while capturing both structural network effects and culturally embedded interaction patterns. Here, we demonstrate how this approach can be used to address a substantive question in social scientific theory: whether core dysfunctions of social media can be mitigated through platform design.


Social simulation has long provided a way to explore counterfactuals. ABMs model how micro-level interactions generate macro-level outcomes \cite{epstein1996growing, gilbert2005simulation, miller2009complex} and have been widely applied to study online polarization, misinformation diffusion, and echo chambers \cite{flache2017models, bruch2015agent, tornberg2018echo, tornberg2021modeling, banisch2019opinion}. However, conventional ABMs often rely on simple decision rules, limiting their ability to represent reasoning, interpretive processes, and dialogue \cite{byrne2022complexity, marres2017digital}. This constrains their usefulness for phenomena -- like online political discourse -- at the intersection of cultural and structural dynamics \cite{pachucki2010cultural}. LLM-based agents address these limitations by enabling nuanced, conversational, and contextually grounded interactions \cite{bail2024can, park2023generative, tornberg2023simulating}. They combine the interpretive richness of language models with the capacity of ABMs to explore emergent dynamics, offering a new means to investigate how design interventions might shape online environments \cite{gu2025large, gao2023s, he2023homophily, mou2024unveilingtruthfacilitatingchange, yang2024oasis, liu2024skepticism, liu2024tiny}.


In this study, we use generative social simulation to test prosocial interventions on a minimal social media platform. In our model, LLM-based agents -- each with a distinct persona -- can post, repost, and follow others. Despite its simplicity, the following-network resulting from the social media model  reliably reproduces three widely discussed pathologies: (1) \emph{echo chambers} formed through homophilous ties, (2) extreme concentration of visibility among a small elite, and (3) a \emph{social media prism} \cite{bail2022breaking} in which politically extreme users hold disproportionate influence. We then implement six platform-level interventions drawn from proposals in the literature to promote prosocial outcomes, ranging from bridging algorithms to hiding engagement metrics. The results are sobering: improvements are modest, no intervention fully disrupts the mechanisms driving these outcomes, and some changes worsen the problems they aim to solve. These patterns suggest that such dynamics may be structurally embedded in the architecture of social media, raising the possibility that meaningful reform will require more fundamental redesign.


\section{Problems of Social Media}
\label{sec:problems}



While the role of social media in relation to rising polarization \cite{tucker2018social}, radicalization \cite{tornberg2024intimate}, misinformation \cite{guess2019less}, and political disengagement \cite{boulianne2015social} remains subject to substantial academic debate, there is widespread agreement that platforms often fail to afford the communicative conditions needed for constructive democratic deliberation and civic engagement \cite{tucker2018social}. Three interrelated platform mechanisms in particular have been highlighted as structural impediments to constructive political debate. 

First, social media enable users to fragment and engage in selective exposure, forming ideologically homogeneous ``echo chambers'' or ``filter bubbles'' \cite{sunstein2018republic,pariser2011filter,bakshy2015exposure,conover2011political}. While the ubiquity of echo chambers remains contested \cite{cinelli2021echo, terren2021echo, garimella2018political, conover2011political, quattrociocchi2016echo, caetano2018using, bakshy2015exposure, kang2012using, de2011tie, 10.1145/2180861.2180866, 10.1145/2645710.2645734, guess2018avoiding,dubois2018echo,tornberg2024intimate}, there is widespread agreement that exposure to diverse viewpoints is a necessary (if not sufficient \cite{bail2018exposure}) condition for democratic deliberation \cite{dryzek2002deliberative,benson2021epistemic}.

Second, platform algorithms -- optimized to maximize user engagement -- often have the unintended effect of amplifying outrage, conflict, and sensationalism \cite{brady2021social,berry2013outrage}. Empirical studies show that negative and moral-emotional language is more likely to go viral \cite{brady2017emotion}. As Bail \cite{bail2022breaking} argues, this creates a ``social media prism'' that distorts political perceptions and deepens divisions. Such dynamics undermine the conditions for deliberation, as constructive debate requires communicative (rather than strategic) action, and norms of mutual respect and justification \cite{habermas1984theory, mansbridge2010deliberative}.

Third, social media platforms reproduce and often intensify inequalities in visibility, voice, and influence. While early scholarship assumed that platforms would level the playing field \cite{castells2009communication}, digital participation is governed by winner-take-all dynamics characteristic of attention economies \cite{tornberg2022sharing,van2013culture}. A small number of highly visible users and accounts command the vast majority of attention and influence \cite{myers2014information,zhu2016attentioninequalitysocialmedia}, while most users remain peripheral. These structural asymmetries again undermine deliberative democracy, which requires that all participants have an equal opportunity to contribute to and shape public discourse \cite{fraser1990rethinking,landemore2020open,young2000inclusion}.

While scholarly debates on the impact of social media on political life remain ongoing \cite{tucker2018social}, substantial evidence suggest that these three structural limitations -- fragmentation, amplification of conflict, and inequality of influence -- undermine the conditions for constructive deliberation and are evident, to varying degrees, across platforms \cite{habermas1996between, cohen1989deliberative, mansbridge2012place}.

In response to these challenges, a growing body of research has suggested focusing on the aim of designing more \textit{prosocial} platforms -- that is, platforms that better afford constructive, inclusive, and respectful forms of exchange \cite{dorr2025research,bak2025moving}. We now turn to outlining our model that seeks to test a series of suggested interventions aimed at achieving these aims.  



\section{Model Description}
Generative social simulations offer a novel framework for studying complex social dynamics, allowing for richer representations of human behavior than conventional ABMs \cite{park2023generative, chan2023chateval, xiao2023simulating, li2023large, aher2023using, xu2023exploring, mandi2024roco}. Building on the view of social media as a sociotechnical system shaped by structural and cultural feedback loops \cite{bak2025moving}, we adopt a complex systems approach to model design. Our goal is not to recreate a fully realistic social media ecosystem or precisely match empirical distributions, but to construct a minimal environment capable of reproducing well-documented macro-level patterns (e.g., homophily, attention inequality, partisan amplification). This allows us to focus on identifying and testing the mechanisms that may underlie these patterns. Following the tradition of minimal modeling \cite{axelrod1997advancing}, we prioritize capturing general mechanisms over fine-tuned calibration \cite{larooij2025large}, while ensuring that the emergent dynamics remain plausible and consistent with stylized empirical observations.


Our goal is to simulate a stylized social media environment to assess whether it reproduces key dysfunctions identified in the literature -- such as polarization, attention inequality, and engagement-driven distortion -- and whether these are mitigated by prosocial interventions. The model centers on a population of simulated users, each represented by a persona drawn from the American National Election Studies (ANES) dataset \cite{ANES2020}. These personas reflect real-world distributions of age, gender, income, education, partisanship, ideology, religion, and personal interests. We extend them using an LLM to generate richer user biographies, including inferred occupations and detailed hobbies, which serve as user profiles within the simulation (see Supplementary Material). 

\begin{figure}[t]
    \centering
    \includegraphics[width=\textwidth]{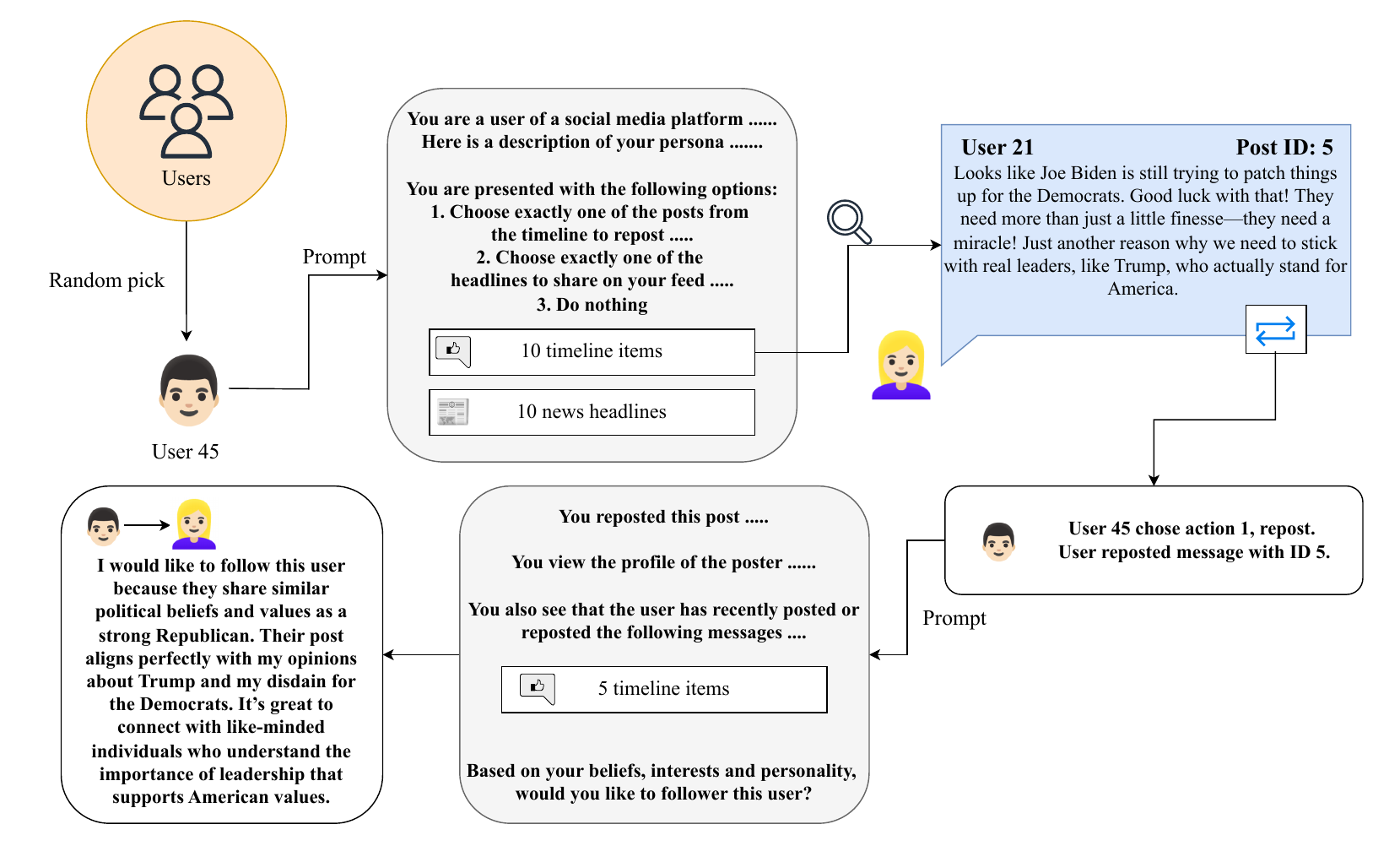}
    \caption{Example of one simulation round.}
    \label{figsim}
\end{figure}

Agents interact asynchronously in discrete time steps (see Fig~\ref{figsim}).  At each step, a randomly selected user may write a new post in response to a news item, repost existing content, or follow another user. Timelines consist of ten posts: five from followed users and five drawn from high-engagement content posted by non-followed users, with repost probability used as a proxy for algorithmic amplification. Content selection, reposting behavior, and user follow decisions are guided by LLM-generated responses to natural language prompts incorporating user biographies, recent posts, and news content. The news feed is populated from a dataset of 210,000 news items \cite{misra2021sculpting, misra2022news}, with a random subset of ten headlines presented to each user considering a new post. 

In the main analysis, GPT‑4o‑mini was used to model users. We however also replicated the base analyses with llama‑3.2‑8b and DeepSeek‑R1, resulting the same qualitative patterns (see Supplementary Material). Further details on simulation architecture, prompt design, and decision flows are provided in the Supplementary Material.


We focus on the structure of the resulting social network, examining whether it reproduces the problematic aspects of social media identified in the previous section: 1) political homophily, 2) disproportional influence of more extreme users, and 3) inequality of follower and engagement.

We test six interventions, each grounded in prior scholarship on mitigating the structural problems of social media:

\begin{enumerate}
\item \textbf{Chronological}: Removes algorithmic recommendations so that non-followed posts appear in reverse-chronological order. Prior work shows that chronological or randomized feeds can reduce exposure to polarizing content and yield a more equal distribution of attention \cite{ribeiro2020auditing, bandy2021problematic}. Bandy and Diakopoulos \cite{bandy2021curating}, for example, found that Twitter’s chronological feed disseminated less low-quality news than its algorithmic counterpart.
\item \textbf{Downplay Dominant}: Inverts engagement weighting to reduce the visibility of highly reposted content. This addresses concerns that engagement-optimized algorithms disproportionately amplify sensational or divisive posts \cite{brady2021social, schone2023negative}.
\item \textbf{Boost Out-Partisan}: Increases the visibility of posts from users with opposing political views, scaled by partisan distance. Such ``viewpoint diversification'' strategies have been proposed to broaden exposure to cross-cutting perspectives and reduce ideological segregation \cite{vendeville2022opening, garimella2017balancing}.
\item \textbf{Bridging Attributes}: Prioritizes posts with high scores on empathy- and reasoning-related attributes, using Perspective API’s Bridging Attributes \cite{saltz2024rerankingnewscommentsconstructiveness}. These “bridging algorithms” aim to elevate content that fosters mutual understanding and deliberative norms over emotional provocation or ideological extremity \cite{bail2022breaking, ovadya2023bridging, wojcik2022birdwatch, kolhatkar2017constructive}.
\item \textbf{Hide Social Statistics}: Obscures repost and follower counts to reduce social influence cues. Engagement metrics have been identified as drivers of inequality of attention, vulnerability to misinformation \cite{Avram_2020}, and amplification of outrage \cite{brady2021social}. Removing these cues has been proposed as a way to dampen such effects.
\item \textbf{Hide Biography}: Removes user biographies from follow prompts, limiting exposure to identity-based signals. Obscuring such cues may reduce echo chamber formation and limit the spread of disinformation \cite{diaz2023disinformation}.
\end{enumerate}

In all recommender interventions (1--4), the curated timeline includes five posts from followed users and five from non-followed users, with only the latter subject to intervention. Full implementation details are provided in Supplementary Material.

\section{Results}\label{sec2}

\begin{table}[h]
\caption{Summary statistics from five runs of the base simulation model with 500 users over 10,000 steps using GPT-4o-mini (for results with other models, see SI). The table reports key network and behavioral metrics: E–I index (measuring homophily), correlation between partisanship and number of followers, correlation between partisanship and repost activity, and Gini coefficients for the distributions of followers and reposts.}\label{tab1}
\begin{tabular*}{\textwidth}{@{\extracolsep\fill}lccccc}
\toprule%
 & E-I Index & \makecell{Corr.\\ partisan - followers} & \makecell{Corr.\\ partisan - reposts} & \makecell{Gini\\ followers} & \makecell{Gini\\ reposts}\\
\midrule
Run 1  & -0.74 & 0.01 & 0.03 & 0.83 & 0.94 \\
Run 2  & -0.86 & 0.12  & 0.11  & 0.83 & 0.94 \\
Run 3  & -0.85 & 0.14  & 0.10  & 0.82 & 0.94 \\
Run 4  & -0.89 & 0.14  & 0.12  & 0.82 & 0.93 \\
Run 5  & -0.84 & 0.14  & 0.07  & 0.84 & 0.95 \\
\midrule
\textbf{Avg.}  & \textbf{-0.84} &\textbf{ 0.11}  & \textbf{0.09 } &\textbf{ 0.83} &\textbf{ 0.94} \\
\botrule
\end{tabular*}
\end{table}

\begin{figure}[h]
\centering
\begin{subfigure}[b]{0.45\textwidth}
    \centering
    \includegraphics[width=\textwidth]{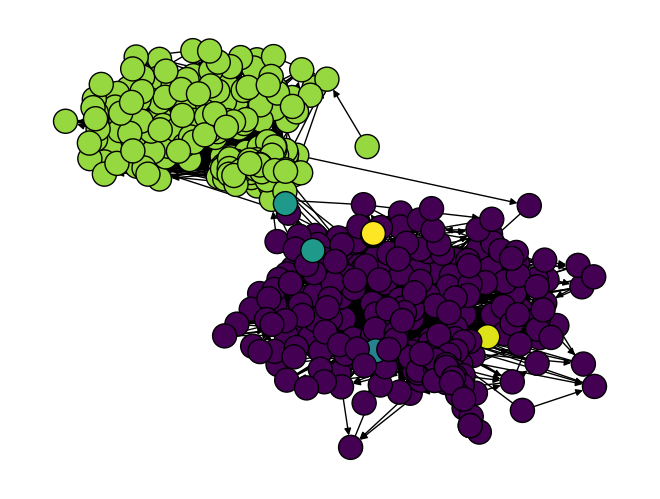}
    \caption{Communities detected by the label-propagation algorithm in a simulated follower network. Each node represents a user, and edges indicate directed follower relationships. Node colors represent algorithmically identified communities.}
    \label{fig2:sub1}
\end{subfigure}
\hfill
\begin{subfigure}[b]{0.45\textwidth}
    \centering
    \includegraphics[width=\textwidth]{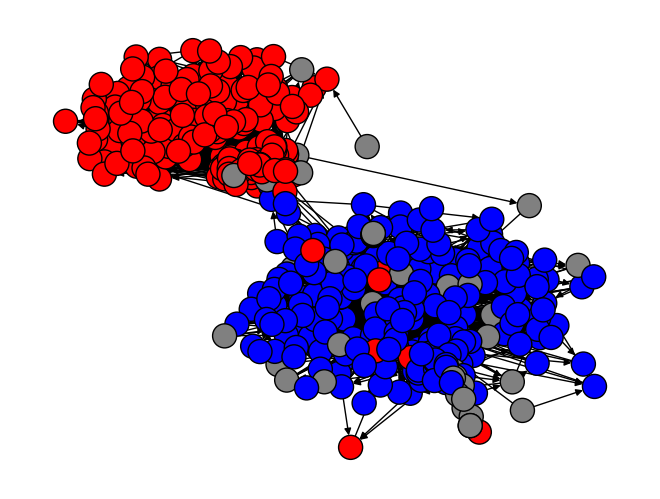}
    \caption{Same network layout as in (a), but with node colors indicating users' political affiliation: red for Republican, blue for Democrat, and gray for non-partisan.}
    \label{fig2:sub2}
\end{subfigure}
\caption{Community structure in the simulated network. Panel (a) shows clusters identified through label-propagation, while panel (b) maps these clusters onto users' political affiliations, revealing partisan segregation.}
\label{fig2}
\end{figure}

\noindent We begin by examining the outcomes of the base platform. Strikingly, despite its simplicity, we find that the model reproduces three key features commonly associated with online political platforms: ideological homophily, unequal attention distribution, and preferential engagement with polarizing content.

First, agents spontaneously form homogeneous communities, with follower ties heavily skewed toward co-partisanship. Across five runs, the average E-I index is $-0.84$ (Table~\ref{tab1}), indicating a strong preference for intra-partisan connections. Community detection via label propagation confirms this pattern: clusters identified purely from network structure (Figure~\ref{fig2:sub1}) closely align with political affiliation (Figure~\ref{fig2:sub2}). 

Secondly, the simulation also produces a highly unequal distribution of visibility and influence. The average Gini coefficient for followers is 0.83, with the top 10\% of users accounting for approximately 75–80\% of all followers. Inequality is even more pronounced in content amplification: reposts exhibit a Gini coefficient of 0.94, with 10\% of posts receiving 90\% of all reposts, while the vast majority receive none (Table~\ref{tab1}). This is in line with a preferential attachment dynamics, in which attention attracts attention \cite{barabasi1999emergence}. 

Finally, we observe correlations between political extremity and engagement. Users with more partisan profiles tend to receive slightly more followers ($r = 0.11$) and reposts ($r = 0.09$). While relatively weak, this correlation suggests the presence of a ``social media prism,'' \cite{bail2022breaking} where more polarized users and content attract disproportionate attention.

Taken together, these results demonstrate that even a minimal platform with posting, reposting, and following -- absent complex recommendation algorithms -- can reproduce core dysfunctions of real-world platforms. These baseline dynamics moreover provide a critical testbed for evaluating whether our interventions can shift user behavior and network-level patterns. 

\subsection*{The Effect of Interventions}
\begin{table}[h]
\caption{Descriptive results for the six interventions compared to the base model. Measures represent averages over five simulation runs.}\label{tab2}
\begin{tabular*}{\textwidth}{@{\extracolsep\fill}lccccccc}
\toprule%
 & \makecell{Action\\ Repost} & \makecell{Action\\ Post} & \makecell{Follow} & \makecell{Max.\\ Followers} & \makecell{Avg.\\ Followers} & \makecell{Max.\\ Reposts} & \makecell{Avg.\\ Reposts} \\
\midrule
\textbf{Base model}  & \textbf{52.5\%} &\textbf{47.4\%}  & \textbf{73.6\%} & \textbf{203.4} & \textbf{6.9} & \textbf{243.2} & \textbf{1.1} \\
\midrule
Chronological  & 53.9\% & 46.1\%  & 69.5\% & 56 & 6.9 & 57.2 & 1.1 \\
Downplay Dominant  & 55.2\% & 44.8\%  & 74.2\%  & 132.2 & 7.4 & 121.4 & 1.2 \\
Other Partisan  & 51.6\% & 48.3\%  & 77.1\%  & 188.0 & 6.9 & 181.2 & 1.1 \\
Bridging Attributes  & 51.4\% & 48.6\%  & 64.3\%  & 168.2 & 5.9 & 180.4 & 1.1 \\
Hide Social Statistics & 58.6\% & 41.4\% & 81.5\% & 189.8 & 8.4 & 169.4 & 1.4 \\
Hide Biography & 49.6\% & 50.4\% & 68.5\% & 192.4 & 6.1 & 199.6 & 1.0 \\
\botrule
\end{tabular*}
\end{table}

\begin{figure}[t]
    \centering
    \includegraphics[width=0.7\textwidth]{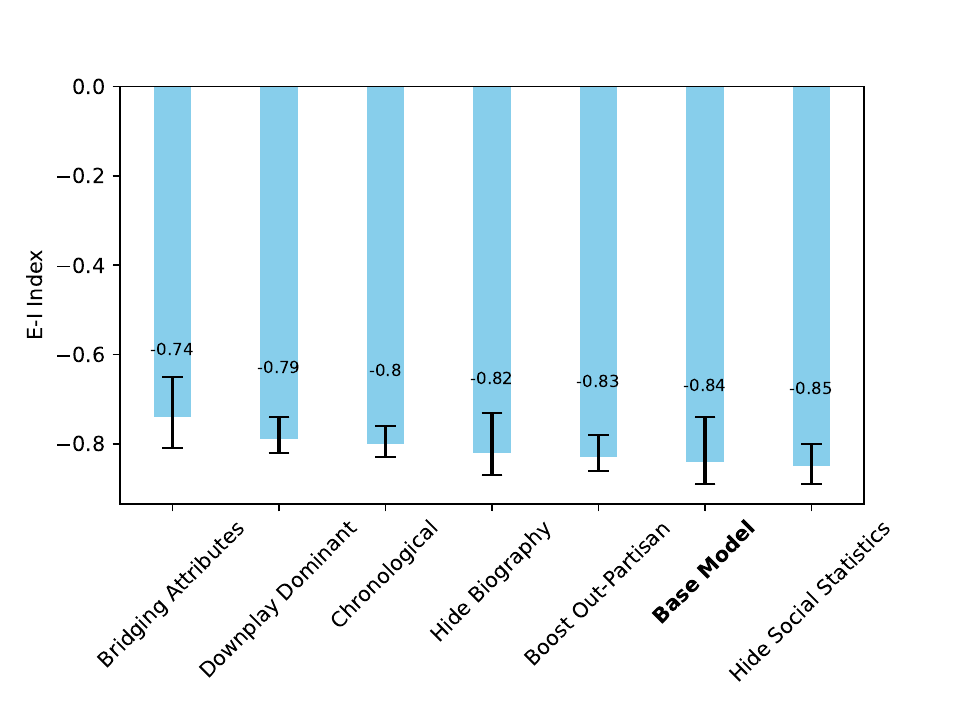}
    \caption{Average E–I index (external–internal index) for each intervention, based on five independent simulation runs. The E–I index measures the relative proportion of cross-group (external) versus within-group (internal) ties in the follower network, with values closer to 0 indicating more cross-partisan connections and values closer to -1 indicating stronger partisan homophily. Error bars show ±1 standard error across runs. Among the tested interventions, \emph{Bridging Attributes} produced the largest reduction in homophily (–0.74 vs. –0.84 in the base model), though all conditions retained a strong preference for following co-partisans.}
    \label{fig3}
\end{figure}

\begin{figure}[t]
    \centering
    \includegraphics[width=0.7\textwidth]{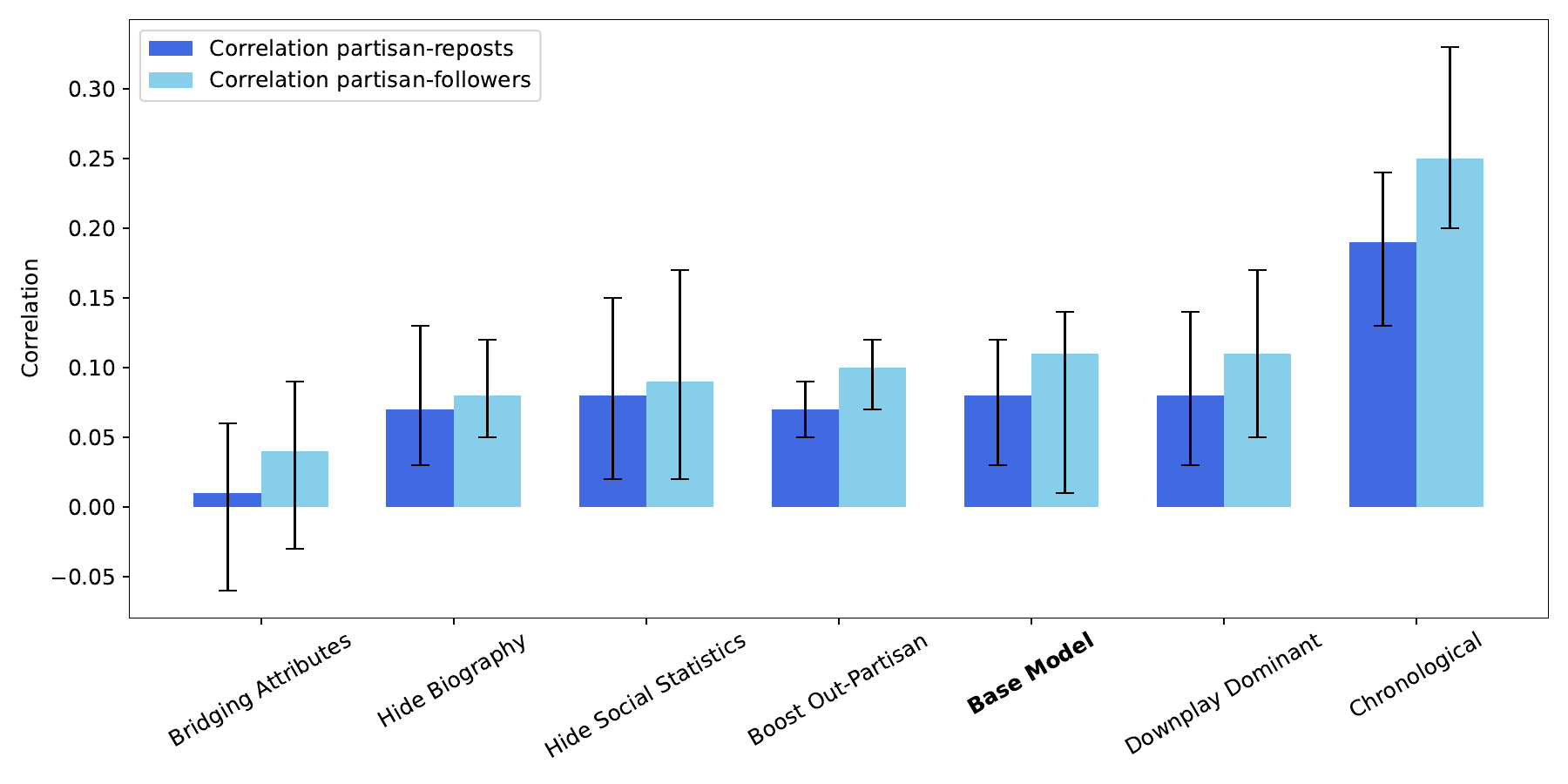}
    \caption{Average Pearson correlation between a user's partisan extremity and their number of followers (light blue) or reposts received (dark blue), for each intervention, averaged over five simulation runs. Positive correlations indicate that more partisan users tend to attract greater visibility and engagement, consistent with the ``social media prism'' effect. Error bars show ±1 standard error across runs. The \emph{Bridging Attributes} intervention substantially reduced these correlations relative to the base model, while \emph{Chronological} ordering increased them.}
    \label{fig4}
\end{figure}

\begin{figure}[t]
    \centering
    \includegraphics[width=0.7\textwidth]{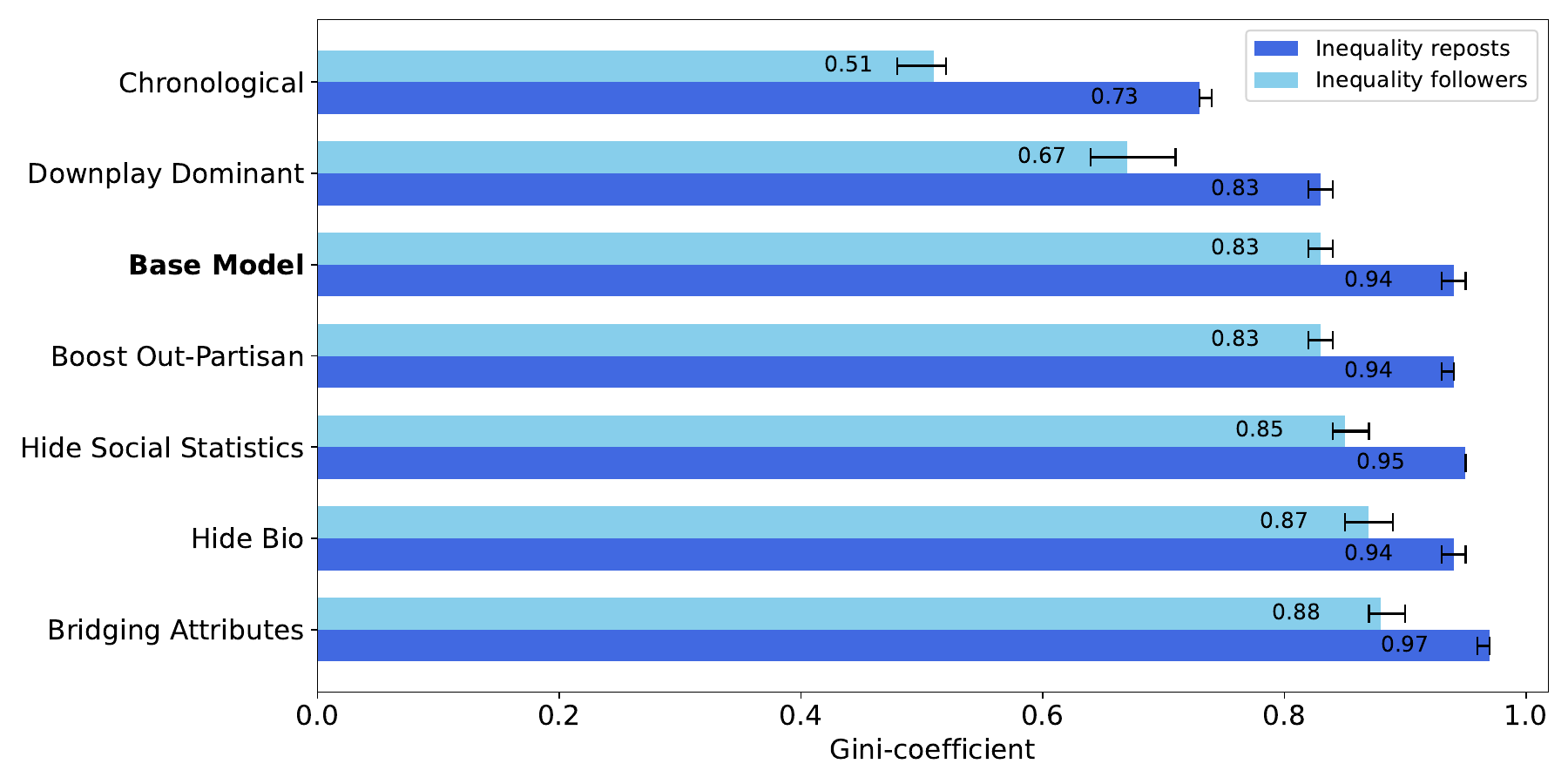}
    \caption{Average Gini coefficient of the distribution of followers (light blue) and reposts (dark blue) for each intervention, averaged over five simulation runs. Higher values indicate greater concentration of attention among a small number of users or posts, with 1 representing complete inequality. Error bars show ±1 standard error across runs. The \emph{Chronological} intervention yielded the lowest inequality in both followers and reposts (Gini = 0.51 and 0.73, respectively), indicating a substantial flattening of the attention distribution. Most other interventions had relatively small effects on inequality compared to the base model, with \emph{Hide Biography} and \emph{Bridging Attributes} showing slightly higher concentration than baseline.}
    \label{fig5}
\end{figure}

We next evaluate the six interventions proposed to address key dysfunctions of social media platforms, comparing their outcomes to the base model. We implemented each intervention in an idealized form -- more extreme than what would be plausible in a commercial platform -- to test its maximum potential effect under controlled conditions. This approach allows us to treat observed changes as upper bounds on real-world impact.

Table~\ref{tab2} summarizes behavioral patterns across conditions, while Figures~\ref{fig3}–\ref{fig5} report effects on political homophily (E-I index), attention inequality (Gini coefficients), and partisan engagement patterns.

\textit{Chronological ordering} -- removing engagement-based ranking -- had the strongest effect on reducing attention inequality. While the overall rates of posting, reposting, and following remained similar to the baseline model, the concentration of followers dropped sharply, and the concentration of reposts also declined (Figure~\ref{fig5}). The reason for this effect is that the intervention effectively breaks the feedback loop between post visibility and popularity. However, the intervention did not reduce ideological homophily, and moreover comes at a cost. First, prior work indicates that chronological feeds often reduce user engagement, raising concerns about viability \cite{bandy2021curating, guess2023social}. Second, the intervention also had a surprising negative side-effect: it intensified the correlation between political extremism and influence, thus further warping the social media prism (Figure~\ref{fig4}). This may be a consequence of more extreme content standing out more sharply against a neutral backdrop in the absence of algorithmic filtering. 

\textit{Downplaying dominant voices} -- prioritizing posts with fewer reposts -- also reduced inequality, albeit to a lesser extent. It lowered maximum follower and repost counts (Table~\ref{tab2}) and reduced Gini coefficients, but had no measurable effect on partisan amplification or homophily.

\textit{Boosting out-partisan content} had little impact across any outcome dimension. Despite increased exposure to ideologically distant posts, users continued to engage primarily with like-minded content, echoing findings that cross-partisan exposure is insufficient for promoting bridge-building \cite{bail2018exposure, bright2022individuals, yang2020mitigating}.

\textit{Bridging attributes}, designed to promote high-quality, constructive content, had more nuanced effects. It was the only intervention to substantially weaken the link between partisanship and engagement (Figure~\ref{fig4}) and modestly increased cross-partisan connections (Figure~\ref{fig3}). However, it also increased inequality: visibility became concentrated among a narrow set of high-scoring posts, highlighting a trade-off between content quality and representational diversity (Figure~\ref{fig5}).

\textit{Hiding social statistics} and \textit{hiding biographies} had minimal effect on the structural dynamics of the network. Homophily, inequality, and partisan amplification remained largely unchanged. However, hiding social statistics did lead to a modest increase in follow and repost behavior (Table~\ref{tab2}), suggesting that users rely on such cues to assess social value and reach. Motivations generated by the agents indicated that users often used follower counts to gauge influence, and hiding this information reduced status-based filtering -- though without disrupting underlying partisan preferences.


\section{Discussion \& Conclusion}\label{sec3}
This study has demonstrated that key dysfunctions of social media -- ideological homophily, attention inequality, and the amplification of extreme voices -- can arise even in a minimal simulated environment that includes only posting, reposting, and following, in the absence of recommendation algorithms or engagement optimization. 

\textit{First}, agents spontaneously formed ideologically homogeneous communities, with follower ties overwhelmingly concentrated within partisan lines. This emergent segregation mirrors real-world patterns of ideological homophily and echo chambers on many platforms, despite the absence of any algorithmic bias or filtering. \textit{Second}, users with stronger partisan identities accrued more followers and reposts. This pattern is consistent with the ``social media prism'' effect, whereby emotionally charged or extreme content and users receive disproportionate visibility \cite{rathje2021out, pandey2023generation, lim2022opinion, whittaker2021recommender}. Although the effect was modest, its emergence in a minimal environment suggests that such dynamics are not necessarily algorithm-induced, but can be self-reinforcing under basic conditions of selective engagement. \textit{Third}, attention was highly unequally distributed: a small subset of users and posts attracted the vast majority of followers and reposts, replicating the power-law dynamics and elite concentration observed on real-world platforms \cite{zhu2016attentioninequalitysocialmedia, myers2014information, aparicio2015model}. This pattern reinforces prior work showing that power-law distributions are robust social regularities that emerge across a wide range of systems \cite{barabasi1999emergence}.

The emergence of these properties from a minimal platform suggests that these problems may be rooted not in the details of platform implementation or algorithms, but in deeper structural mechanisms: they stem from the entangled dynamics of content engagement and network formation. Reposting does not merely amplify content: it incrementally constructs the follower network, as users are exposed to others via reposts from accounts they already follow. Centrally, this means that the affective, reactive, and partisan nature of reposting decisions \cite{rathje2021out, bail2018exposure, bright2022individuals} directly determines who becomes visible and who gains followers. This creates a self-reinforcing cycle: affective engagement drives network growth, which in turn shapes future exposure. These dynamics feed back into content visibility, reinforcing ideological homogeneity, attention inequality, and over-representation of extreme users and content.

We furthermore evaluated six widely discussed interventions intended to promote more prosocial online environments. These should be taken as upper-bound interventions, in the sense that they are more extreme than what could plausibly be implemented in a real-world platform, and that the impact on the user experience was not considered. While several showed moderate positive effects, none fully addressed the core pathologies, and improvements in one dimension often came at the cost of worsening another. 


Taken together, our findings challenge the common view that social media's dysfunctions are primarily the result of algorithmic curation. Instead, these problems may be rooted in the very architecture of social media platforms: networks that grow through emotionally reactive sharing. If so, improving online discourse will require more than technical tweaks -- it will demand rethinking the fundamental dynamics of interaction and visibility that define these environments.

This work also marks one of the first uses of generative social simulation to contribute to social scientific theory. Importantly, the central mechanism identified by this simulation -- the feedback between reactive engagement and network formation -- would have been challenging to capture without the use of generative social simulation, as they include both structural and cultural dimensions \cite{pachucki2010cultural}. Positioned between empirical analysis and abstract modeling, generative simulation offers a powerful and flexible approach for studying emergent social dynamics. Yet it also raises important questions around realism, interpretability, and validity \cite{larooij2025large}. LLM-based agents, while offering rich representations of human behavior, function as black boxes and carry risks of embedded bias. The findings of this study should hence not be taken as definitive conclusions, but as a starting-point for further inquiry. 

Several limitations warrant mention. First, our model does not capture the user experience, a critical factor for real-world platform viability. The key question of whether prosocial design can coexist with high engagement and user satisfaction remains unanswered. Second, validation poses a persistent challenge. Generative simulations are even harder to calibrate to empirical data than conventional ABMs \cite{larooij2025large}, and LLM-based agents introduce additional complexities, including hallucination, limited controllability, and embedded social biases \cite{bolukbasi2016man, wan2023kelly}. While our complex systems approach prioritizes emergent patterns over precise behavioral fidelity \cite{bak2025moving}, further research is needed to determine when and how generative simulations can yield reliable social scientific insights. Finally, the approach is computationally intensive: simulating 500 agents over 10{,}000 steps required several hours per run, constraining our ability to systematically explore the parameter space. Scaling to more complex environments will demand innovation in both simulation design and computational infrastructure.

\section{Code Availability Statement}
The full code used for this paper is available on GitHub: \url{https://github.com/cssmodels/prosocialinterventions}

\bibliography{sn-bibliography}

\end{document}